# Middleware Implementation in Cloud-MANET Mobility Model for Internet of Smart Devices

**Tanweer Alam**,

*tanweer03@iu.edu.sa*

Faculty of Computer and Information Systems, Islamic University in Madinah, Saudi Arabia

**Abstract**
The smart devices are extremely useful devices that are making our lives easier than before. A smart device is facilitated us to establish a connection with another smart device in a wireless network with a decentralized approach. The mobile ad hoc network (MANET) is a novel methodology that discovers neighborhood devices and establishes connection among them without centralized infrastructure. Cloud provides service to the MANET users to access cloud and communicates with another MANET users. In this article, I integrated MANET and cloud together and formed a new mobility model named Cloud-MANET. In this mobility model, if one smart device of MANET is able to connect to the internet then all smart devices are enabled to use cloud service and can be interacted with another smart device in the Cloud-MANET framework. A middleware acts as an interface between MANET and cloud. The objective of this article is to implement a middleware in Cloud-MANET mobility model for communication on internet of smart devices.

## 1. Introduction

The smart device to smart device communication in the cloud-MANET framework is a novel methodology that discovers and connected nearby smart devices with no centralized infrastructure. The existing cellular network doesn't allow to connect all smart devices without centralized infrastructure even if they are very near to each other. The proposed technique will be very useful in machine to machine (M2M) networks because, in M2M network, there are several devices nearby to each other. So the implementation of MANET model in the smart device to smart device communication can be very efficient and useful to save power as well as the efficiency of spectrums. The cloud-based services in MANET modeling for the device to device communication can be a very useful approach to enhance the capabilities of smart devices. The smart device users will use cloud service to discover the devices, minimize useful information in a big data and can process videos, images, text, and audio. In this article, I proposed a new middleware framework to enhance the capability of MANET and cloud computing on the internet of smart devices that can be useful in the 5G heterogeneous network. In proposed framework, the smart device will consider as service nodes. We consider Android framework to implement proposed idea. Android operating system is more popular than another operating system in the world. Android OS is a freely available platform for cell phones and is produced by individuals from the Open Handset Alliance.

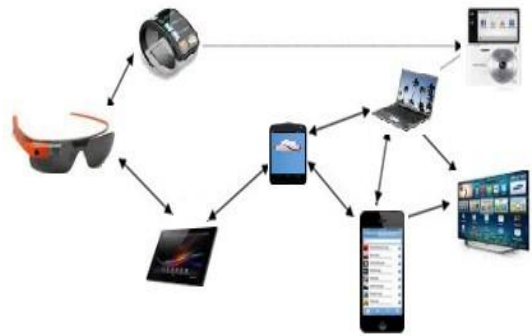

Fig. 1. MANET of smart devices

The Open Handset Alliance is a gathering of more than 40 companies, including Google, ASUS, Garmin, and HTC

and so on. These organizations have met up to quicken and enhance the advancement of cell phones. Presently android smart devices are increasing exponentially in the world [9]. These smart devices have strong multimedia features. These features are helpful too much to the android user. The android users use these features to share multimedia quickly. Many android clients are increasing their interest to share videos and take photos by embedded camera. Also, the popularity of android devices is increasing in the developing projects of entertainment. In this project, clients upload their latest photographs to the system [38]. Presently, the android smart devices are very popular with the addition of its high capability. The newly feature of Android devices is Wi-Fi Direct [44]. Using this feature the wireless technologies [29] provide support to their users to make the very good use of ad hoc network with smart devices at all time and everywhere [6]. The MANET is a decentralized network [28] that created by the wireless devices with infrastructure-less environment [47].

The communication between devices in ad hoc environment is to be unique [12] and innovative [33]. The intercommunication without centralized approach is a very powerful mechanism [43] that provides secure communication to the users [37]. The ad hoc network communication of android smart devices can play a most important role when the cellular network fails. Cloud computing has been regarded as one of the most popularized computing paradigms. Cloud computing gives its customers with three essential administration models: SaaS, PaaS, and IaaS. Software as a service (SaaS) is mainly intended to end users who need to use software as a part of their daily activities.

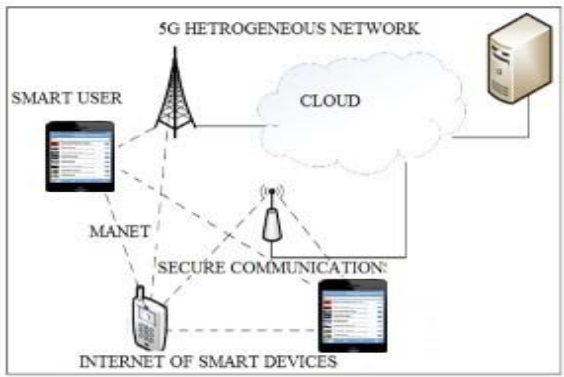

Fig. 2.Cloud-MANET integration model

Platform as a service (PaaS) is mainly intended for application developers who need platforms to develop their software or application. Infrastructure as a service (IaaS) is mainly intended to network architects who need infrastructure capabilities. The communication security challenges and threats for communicating in cloud perspective internet of smart devices are the most important aspect. The smart device to smart device communication in the cloud-MANET framework of the internet of things is a novel methodology that discovers and connected nearby smart devices with no centralized infrastructure. The smart devices will connect in the range of Wi-Fi wireless network [5]. All Android smart devices within the range can communicate with each other without cellular network [15]. I mean communication in own created network. This own created network is the special network without centralized approach i.e. Ad Hoc Network [17]. The figure1 represents the ad hoc network among some smart devices without cellular networks [3].

The Cloud MANET mobility model is an integration model of Cloud computing and MANET technologies. The functionality of MANET is depended on the mobility of its nodes and connectivity also resources such as storage and energy efficiency [8]. In Cloud computing, cloud providers retain network infrastructure, storage facilities, and software applications that support flexibility, efficiency, and scalability [41].

In Cloud MANET mobility model, smart devices of MANET can communicate with each other but at least one smart device must be connected with cellular or Wi-Fi networks. All smart devices of MANET should be registered in cloud individually. The proposed model will activate in disconnected mode. When a MANET is activated then cloud services will activate in a real time and provide services to the smart devices of MANET. The smart devices send a request to the cloud for a session of connectivity. Cloud provide the best connection to the smart device. The proposed middleware is designed to access android services in ad hoc environment. It exists between the user and hardware [26]. It connects with applications in ad hoc networks [13]. Users communicate through application among android devices. The connection is established through middleware [16]. The middleware provides facility to create ad hoc network

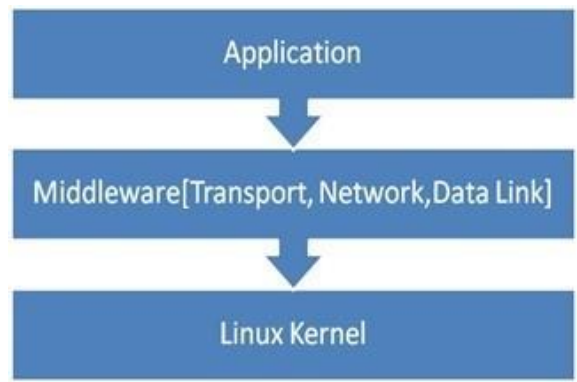

Fig. 3.Middleware between application and Linux kernel

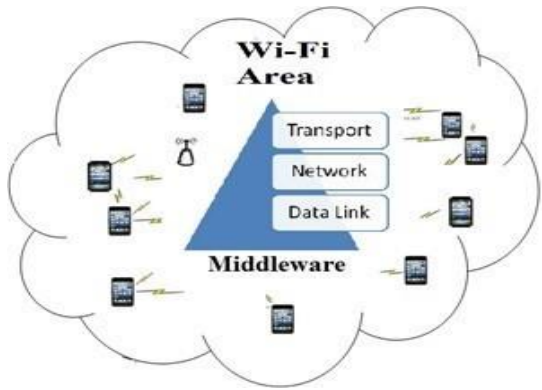

Fig. 4.Middleware among smart devices in Wi-Fi

[24], provide a secure route, transportation and secure connections [27] [20]. It works in the wireless network [10]. So we need Wi-Fi network to establish middleware among android devices.

In proposed middleware framework, the smart devices are exploiting the cloud service in mobile ad hoc network and create a connection among smart devices to communicate each other in the internet of things scenario. These cloud services provide a significant approach for communication in a large number of smart devices using routing protocols. In this article, I focused on implementation of middleware in MANET of smart devices and exploit cloud service by at least one smart device with an internet connection to communicate other smart devices without internet connection within the range of MANET.

## 2. Literature Survey

In the 1980s, with the evolving of the internet, the foundation of an emerging grid computation was established. The foundation involved various principals which employ the internet in a way in which users are provisioned as resource nodes. A grid coordinates these resources nodes and dispenses takes to them thus the entire computation is viewed as a cumulative fashion. The principles paved the way of a novel computing paradigm which eventually carved today's distribution concepts. In the 1990s, the concept of virtualization was driven to the application tier. It followed by employing virtualized private network connections which share the same physical channel. In 1991, Theodore S. Rappaport published an article entitles The wireless revolution, in this paper he presented the wireless communications is the emerging technology as a key for communication among human as well as devices [35]. In 1994, Andy Harter and Andy Hopper published the article entitled A Distributed Location system for the active office, in this paper they presented infrared sensor arrangements using badges for communicating among devices and workstations [18]. In 1994, Tristan Richardson, Frazer Nett, Glenford Mapp, and Andy Hopper presented an article on A ubiquitous, personalized computing environment for all Telephone in an X Window System Environment, in this article they presented X windows systems, X protocol for securing the communication between client and server [36]. In the article [21], authors represented System Software for Ubiquitous Computing for integration of different kinds of network, also create a connection among the devices in different types of network. In 2002 researchers published an article entitled Connecting the Physical World with Pervasive Networks, in this article they address the challenges and opportunities of instrumenting the physical world with pervasive networks of sensor-rich, embedded computation [15]. The cloud computing came as a consequence of the continued development of computing paradigms. The emergence of these technologies has established the appearance of (SaaS) software as a service which states that consumers are not required to purchase the software rather than paying according to their own demand. In the mid of 2006, Amazon achieved a prominent milestone by testing elastic computing cloud (EC2) which initialized the spark of cloud computing in it. However, the term cloud computing was not found until March 2007. The following year brought even more rapid development of the newly emerged paradigm. Furthermore, the cloud computing infrastructure services have widened to include (SaaS) software as a service. In the mid of 2012, oracle cloud has been introduced, where it supports different deployment models. It is provisioned as the first unified collection of it solutions which are under continued developments. Nowadays, typing a cloud computing in any search engine will result in a tremendous result. For example, it would result in more than 139,000,000 matches in Google. In 2009, Evan Welbourne et al published an article entitled Building the Internet of Things Using RFID, in this paper authors presented RFID-based personal object and friend tracking services for the IoT that proposed tools can quickly enable [45]. In 2010, Gerd Kortuem et al. published an article on Smart objects as building blocks for the internet of things, in this article they presented the development of a new flow-based programming paradigm for smart objects and the Internet of Things [22]. In 2011, Ahmed Rahmati et al published an article on Context-Based Network Estimation for Energy-Efficient Ubiquitous Wireless Connectivity, in this article they presented context-based network estimation to leverage the strengths and provide ubiquitous energy efficient wireless connectivity [34]. In the article [44] researchers presented Wi-Fi based sensors for the internet of things, they focused on measurement the range performance. In May 2014, Lihong Jiang et al published an article entitled An IoT-Oriented Data Storage Framework in Cloud Computing Platform, they focused on data storage framework that is not only enabling efficient storing of massive IoT data but also integrating both structured and unstructured data [19]. In the article [39], introduced the IoT ecosystem and key technologies to support IoT communications. In 2016, Maria Rita Palattella et al published an article entitled Internet of Things in the 5G Era: Enablers, Architecture and Business Models, in this article they presented 5G technologies for the IoT, by considering both the technological and standardization aspects [30].

## 3. Problem and Research Questions

There is various software in the market to provide connectivity among peoples and smart devices using cloud service and internet. But this software required internet connection always on their smart devices. The Internet is a part of our daily life but sometimes we face problem for network connection, slow speed or no network. Also in a disaster situation, emergency or military rescue operation etc., in that situation people can't access their internet connection to communicate their neighbors. We can't get our information on the site and communicate with our neighbors or world without network bandwidth. Wi-Fi direct was launched in 2010 for communication among nearby devices. It has various features including discovering the neighbor devices, social networking, file sharing and disaster recovering etc., But it operates through the battery so that it is a disadvantage of this technology. The following are the open questions that we are addressing in this article.

Question 1: Is Wi-Fi direct sufficient for communication among smart devices?
Question 2: How can we increase the distance of coverage while the transmission power is limited?
Question 3: How can we connect a large number of devices?
Question 4: How can we stable the connection between smart devices through Wi-Fi Direct?
Question 5: Many users found bugs in this technology. How can we remove these bugs?

## 4. Methodology

The proposed middleware implemented between the application layer and Linux kernel. The Android operating system runs on Linux kernel. The middleware has transportation, discover the new devices and find the shortest path and create a connection among all android devices [25]. The middleware provides facility to android based smart devices to create connection and joins self-created network ie. Ad hoc network [25]. It provides a reliable route to forward data in the transportation [48]. The new proposed middleware is also providing a simple interface to easily useful for non-technical users. It supports almost all applications.

The life of connection is described as the probabilistic function [49] as follows.

$$\text{session(life))} = \left( \frac{\int_{life}^{\infty} \left( \left(\frac{1}{2}\right) - \left(\frac{1}{2}\right) erf\left(\log\left(\frac{u}{\mu}\right) \div \sqrt{2}\sigma\right)\right) du}{\left(\frac{1}{2}\right) - \left(\frac{1}{2}\right) erf\left(\log\left(\frac{life}{\mu}\right) \div \sqrt{2}\sigma\right)} \right)$$

The expression in the integral will be 0 if the limit tends to ∞.

$$\text{session(life)} = \mu \cdot e^{\sigma^2} \cdot \left( \frac{1 erf\left(\sigma^2 \mp \log\left(\frac{life}{\mu}\right) \div \sqrt{2}\sigma\right)}{erfc\left(\log\left(\frac{life}{\mu}\right) \div \sqrt{2}\sigma\right)} \right)$$

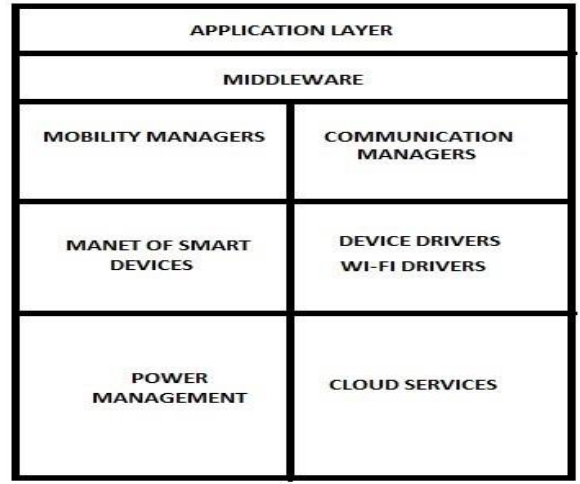

Fig. 5. Middleware architecture

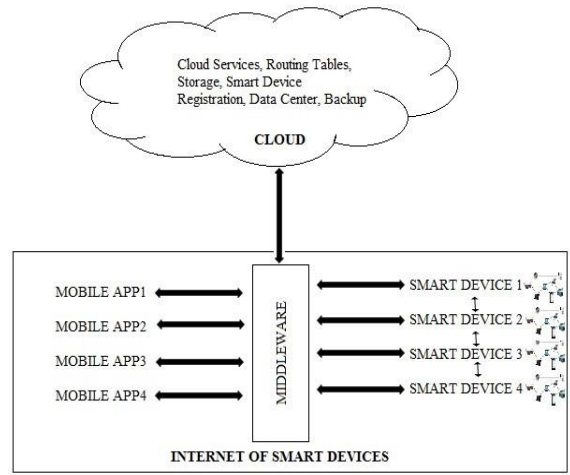

Fig. 6. Middleware implementation in Cloud-MANET integration model

After computing session life by using above probabilistic function, every smart device requires to compute the values of $\sigma$ and $\mu$.

These two parameters are related to the connection establishment among MANETs and Cloud service that can be measured through smart devices using the following function. $e^{\mu(1/2)\sigma^2}$

When a smart device estimate the connection life between MANET and Cloud, it will transfer or receive data securely. The connection will be activated and stability will be high. We consider that every smart device is assured to establish the route between MANET and cloud when they create session in the cloud. The smart devices can move through the maximum speed 20m/s from one location to another location by using Gauss markov mobility model. The following formula is used to calculate the moving speed and direction of the smart device within MANET range [23].

$$\text{Speed}_t = \lambda \text{Speed}_{t-1} + (1-\lambda)\overline{Speed} + \sqrt{(1-\lambda^2)\text{Speed}_{t-1}^G}$$

and

$$Direction_t = \lambda Direction_{t-1} + (1-\lambda)\overline{Direction} + \sqrt{(1-\lambda^2) Direction_{t-1}^G}$$

The $\lambda$ is used as random degree when computing speed as well as direction of smart device in a duration (t). The transmission ($t_s$) of information ($I_k$) among the number of smart devices ($S_n$) can be estimated during the time interval [$t_i, t_{i-1}$]. The smart devices can moved within the MANET and access the cloud service using the multidimensional function ($\varepsilon^k$).

$$\varepsilon k = C^{Sn \times tk} \times Ik$$

where k=0,1,2,3.......∞(+ve).

If smart devices have moved outside the MANET then k will be negative value. Here we consider that the transformation of information happens simultaneous. The bessel's function can be calculated as follows.

$$B_F(S_n \mid \varepsilon_k, I_k, t_k) = 1 - M_Q\left(\sqrt{\sigma}, \sqrt{\frac{S_n}{\alpha}}\right)$$

Where $B_F$ is the Bessel's method and $M_Q$ is the Marcum's method. The σ and α are the parameters used to calculate Marcum's method.

We know that the probability is proportional to the one divide by information.

$$P_k \propto \frac{1}{I_k}$$

The probability density function for transmission is calculated mathematically as follows.

$$P_k(S_n \mid \varepsilon_k, I_k, t_k) = 1 - \int_{-\infty}^{\infty} M_Q\left(\sqrt{\frac{2\gamma^2}{1-\gamma^2} \times \frac{S_n}{t_k}}, \sqrt{\frac{S_n}{1-\gamma^2} \times 2\gamma}\right) \text{ where } \gamma \neq 1$$

Now we have divided the probability density function of all the connections using the entropy per symbol of all connected devices in 3-dimensional directions.

$$H_{\alpha,\delta,\varepsilon} = \sum_{X=0}^{\infty} \sum_{Y=0}^{\infty} \sum_{Z=0}^{\infty} \alpha_k . \beta^{-1} . \varepsilon_2\left(\frac{\delta}{\alpha}, S_k, \rho\right) . \left[P_{X_k} . P_{Y_k} . P_{Z_k} \log_3\left(\frac{1}{P_{X_k}} \frac{1}{P_{Y_k}} \frac{1}{P_{Z_k}}\right)\right]$$

Here $\varepsilon_2\left(\frac{\delta}{\alpha}, S_k, \rho\right)$ is the Chi-Square distribution method that is used here for convergence. Now we will calculate all the probabilities, entropies in each direction and finally we draw the transition matrix from the probabilities of all connected devices as follows.

$$\begin{bmatrix} P_{111} & P_{112} & P_{113} & .... & P_{11K} \\ P_{211} & P_{212} & P_{213} & .... & P_{21K} \\ ...... & ...... & ...... & .... & ...... \\ P_{K11} & P_{K12} & P_{K13} & .... & P_{K1K} \end{bmatrix} . \begin{bmatrix} P_{111} & P_{121} & P_{131} & .... & P_{1K1} \\ P_{211} & P_{221} & P_{231} & .... & P_{2K1} \\ ...... & ...... & ...... & .... & ...... \\ P_{K11} & P_{K21} & P_{K31} & .... & P_{KK1} \end{bmatrix}$$

Now we will find entropy per symbol row-wise said $H_1, H_2, H_3, .......,H_K$ according to above transition matrix. After findings of $H_1, H_2, H_3, .......,H_K$ we will found the whole entropy per symbol of the smart devices.

$$H = H_1.P_1 + H_2.P_2 + H_3.P_3 + .........H_K.P_K.$$

We have calculated the velocities of smart devices using Gauss-Markov Mobility Model in multidimensional area of MANET. We have tested on simulation using 5, 10 and 50 smart devices at 50 m/s and 100 m/s. We got data that are shown in Tabel 1,2 at the time of testing. Middleware needs some java classes to discover, block devices in a particular location [42].

## 5. Implementation

The middleware in the cloud-MANET framework is implemented in Java programming language in the form of android based mobile application. The Android architecture provides built-in tools to android applications for mobile smart devices [7]. It means that the programmers need only to develop an application using the Android operating system and they can run these applications on different smart devices that powered by Android [2]. Android keeps running on Linux under Dalvik VM. Dalvik has an in the nick of time compiler where the byte code put away in memory is ordered to a machine code. Bytecode can be characterized as middle level. JIT compiler peruses the bytecode in numerous segments and accumulates progressively with a specific end goal to run the project quicker. Java performs keeps an eye on distinctive parts of the code and in this manner, the code is gathered just before it is executed [1]. When it is compiled once, it is stored and set to be prepared for later uses. Linux Kernel Android can bolster administrations of the center framework that provides a level of abstraction between the device hardware [46] and it contains all the essential hardware drivers such that front and rear camera, smart keypad, touch screen etc. Also, the kernel handles networking, Wi-Fi and Bluetooth drivers interfacing to peripheral hardware. The android framework is divided into three layers [3]. The first layer is Application Layer. It is designed for ad hoc applications to simplify the components for reuse [21]. By default, the android operating system uses so many core applications like browsers, wireless services, contact list etc. Google provides so many open source applications for developers.

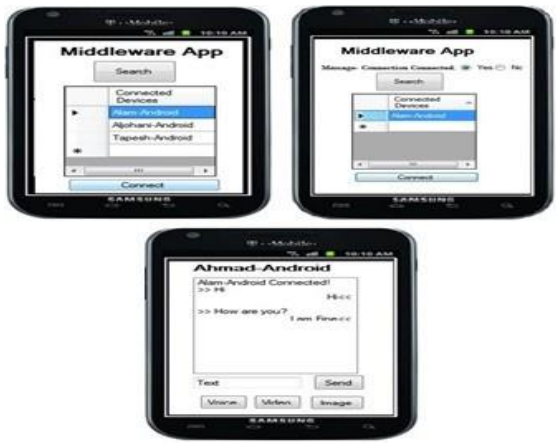

Fig. 7. Middleware implementation using android application

The developer has the possibility to change or modify these applications and make their own applications accordingly. The second layer is libraries and android runtime. In this layer of Android Architecture in ad hoc environment include a group of libraries of different services [11]. The developer can use these services and develop creative functionality in android architecture. This layer provides device manager class, discovery classes of Wi-Fi as well as Bluetooth services. The names of classes are Wi-FiDiscoveryService, Wi-FiBlackListedService, BluetoothDiscoveryService, BluetoothBlackListedService and DeviceManager class. The Wi-FiDiscoveryService class is used to discover all smart devices in the range of Wi-Fi [14]. The Wi-Fi BlackListedService class is used to make a list of all blacklisted smart devices. The BluetoothDiscovery-Service class is used to discover all smart devices in the range of Wi-Fi. The BluetoothBlackListedService class is used to make a list of all blacklisted smart devices [4]. The third layer is Routing and link Layer. In routing layer [31] of Android architecture, in include methods for sending datagram using one of these, unicast, multicast and broadcast in the range of Wi-Fi [32]. This layer also has an event that responsible for notifying of incoming messages. This layer works between network and libraries for discovering. These libraries have discovered methods for discovering immediate neighbors or network contacts [40]. We add the proposed middleware between the application layer and Linux kernel in the android framework with cloud service. We used Wi-FiDiscoveryService class of android in proposed middleware for discovering the smart devices within the range of MANET. I have developed a mobile application for testing middleware in Cloud-MANET model of the internet of smart devices. This application discovers all neighbor smart devices in the range of Wi-Fi using the search button. When we want to connect our smart device to other discovered device, then just click the name in the list box and click connect button. When we click connect button the connection created message will be sent to the appropriate device. When we receive the confirmation from that device we can communicate to each other. Also, we can transfer data, voice, video and image from one device to another device using this android application.

The following procedure should be followed by smart devices.
1. Install the mobile app and register in the cloud. The cloud will provide access permission.
2. Enter smart device id and password to login in the cloud.
3. Store WPA supplicant.conf on every smart device. This leis used to start MANET service on the smart devices. We had connected this le to our developed mobile apps.
4. Start MANET.
5. Searching neighborhood devices within the range of MANET or search through the device id. 6. Click on the searched device and start communication.

In 50 meter range of Wi-Fi, the maximum throughput among smart devices communication was 10 Mbps, 8.1

Mbps, 8.5 Mbps, and 5.8 Mbps for Text, Image, Voice, and Video, respectively [Table.3].

In 200 meter range of Wi-Fi, the maximum throughput among smart devices communication was 10 Mbps, 6.8

TABLE I. TRANSMISSION IN MANET OF SMART DEVICES AT 50 M/S.

| Devices | $\varepsilon_k = 0.1$ | $\varepsilon_k = 0.2$ | $\varepsilon_k = 0.4$ | $\varepsilon_k = 0.6$ | $\varepsilon_k = 0.8$ | $\varepsilon_k = 1$ |
|---|---|---|---|---|---|---|
| 5 | 2 | 2.1 | 2 | 2.2 | 2.5 | 2.4 |
| 10 | 3 | 3.2 | 3.1 | 3.5 | 3.4 | 3.3 |
| 50 | 7 | 7.2 | 7.5 | 7.8 | 7.4 | 7.3 |

TABLE II. TRANSMISSION IN MANET OF SMART DEVICES AT 100 M/S.

| Devices | $\varepsilon_k = 0.1$ | $\varepsilon_k = 0.2$ | $\varepsilon_k = 0.4$ | $\varepsilon_k = 0.6$ | $\varepsilon_k = 0.8$ | $\varepsilon_k = 1$ |
|---|---|---|---|---|---|---|
| 5 | 2.1 | 1.9 | 1.9 | 2.2 | 2.3 | 2.42 |
| 10 | 3 | 2.9 | 3.1 | 2.8 | 2.7 | 3 |
| 50 | 5.1 | 5.5 | 7.5 | 5.3 | 5.2 | 5.6 |

TABLE III. THROUGHPUT IN SMART DEVICES COMMUNICATION IN MANET IN 50-METER RANGE

| Data Types | Mbps |
|---|---|
| Text | 10 |
| Image | 8.1 |
| Voice | 8.5 |
| Video | 5.8 |

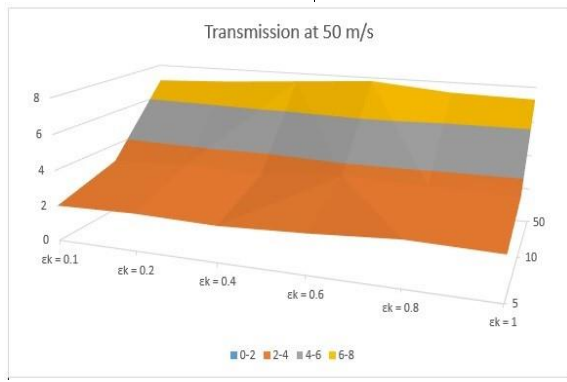

Fig. 8. Transmission in Cloud-MANET at 50 m/s

In 100 meter range of Wi-Fi, the maximum throughput among smart devices communication was 10 Mbps, 7.2 Mbps, 8.5 Mbps, and 4 Mbps for Text, Image, Voice, and Video, respectively [Table.4].

Mbps, 8.2 Mbps, and 2.6 Mbps for Text, Image, Voice, and Video, respectively [Table.5].

Figure. 10 represents the result interpretation of sending the Text, Images, Audio and video files from 50, 100 and 200-meters distance using middleware of cloud-MANET architecture among smart devices.

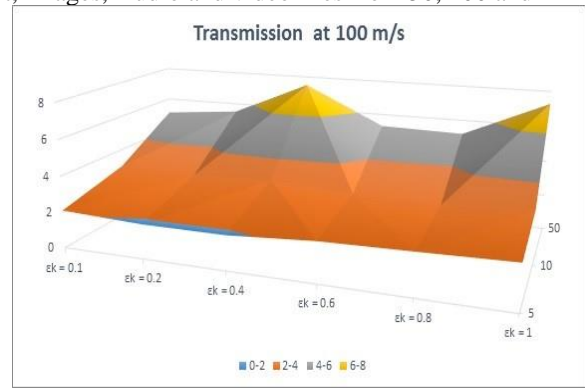

Fig. 9. Transmission in Cloud-MANET at 100 m/s

TABLE IV. THROUGHPUT IN SMART DEVICES COMMUNICATION IN MANET IN 100-METER RANGE

| Data Types | Mbps |
|---|---|
| Text | 10 |
| Image | 7.2 |
| Voice | 8.5 |
| Video | 4 |

TABLE V.        THROUGHPUT IN SMART DEVICES COMMUNICATION IN MANET IN 200-METER RANGE

| Data Types | Mbps |
|---|---|
| Text | 10 |
| Image | 6.8 |
| Voice | 8.2 |
| Video | 2.6 |

## 6. Conclusion

The middleware in Cloud-MANET mobility model is sufficient for communication among smart devices without centralized system while Wi-Fi Direct is not sufficient to establish connection among smart devices using cloud. We can increase the distance of coverage using cloud. The smart device of one MANET is able to connect with another smart device of different MANET using cloud service. We can connect a large number of smart devices together. We can establish connection among smart devices for a long time. There is no bugs in this technology. It is working fine. In the future, we can integrate this technology to internet of things framework.

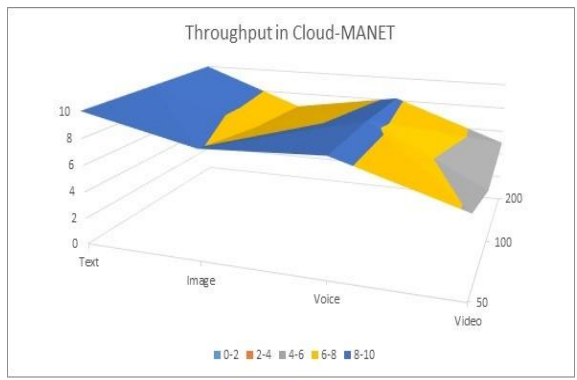

Fig. 10. Testing Text, Image, Voice and Video on middleware in Cloud-MANET Mobility Model

As a consequence of node mobility fixed source/destination paths cannot be maintained for the lifetime of the network. The need for developing middleware technique in a mobile ad hoc network communication for smart devices is to communicate with each other and transfer data, image, voice and video. The android application for connecting smart devices and transfer data in mobile ad hoc environment using cloud service has been done and results were collected in the range of 50 meters, 100 meters, 200 meters respectively. The application has been tested in a Wi-Fi ad-hoc network environment. The results showed successful and expectation for a future scope in the area of mobile ad hoc network and internet of things.